\documentclass[aps,prd,amsmath,amssymb,onecolumn]{revtex4}

\usepackage{graphicx}
\usepackage{dcolumn}
\usepackage{bm}
\usepackage{amsmath}
\usepackage{epstopdf}
\usepackage{amsfonts}
\usepackage{amssymb}
\usepackage{epsfig}
\usepackage{tabularx}
\usepackage{color}
\usepackage{soul}
\usepackage{multirow}
\usepackage{xcolor}

\usepackage[colorlinks=true,citecolor=blue,urlcolor=magenta,breaklinks]{hyperref}
\newcommand{\be}{\begin{equation}}
\newcommand{\en}{\end{equation}}
\newcommand{\bea}{\begin{eqnarray}}
\newcommand{\ena}{\end{eqnarray}}

\newcommand{\bff}{\begin{figure}}
\newcommand{\eff}{\end{figure}}

\begin{document}

\title{Stellar modeling via the Tolman IV solution: The cases of the massive pulsar J0740+6620 and the HESS J1731-347 compact object.}

\author{Grigoris Panotopoulos}

\address{
Departamento de Ciencias F{\'i}sicas, Universidad de la Frontera, Casilla 54-D, 4811186 Temuco, Chile.
\\
\href{mailto:grigorios.panotopoulos@ufrontera.cl}{\nolinkurl{grigorios.panotopoulos@ufrontera.cl}} 
}

\begin{abstract}
We model compact objects of known stellar mass and radius made of isotropic matter within Einstein's gravity. The interior solution describing hydrostatic equilibrium we are using throughout the manuscript corresponds to the Tolman IV exact analytic solution obtained long time ago. The three free parameters of the solutions are determined imposing the matching conditions for objects of known stellar mass and radius. Finally, using well established criteria it is shown that contrary to the Kohler Chao solution, the Tolman IV solution is compatible with all requirements for well behaved and realistic solutions. except for the relativistic adiabatic index that diverges at the surface of the stars. The divergence of the index $\Gamma$ may be resolved including a thin crust assuming a polytropic equation-of-state, which is precisely the case seen in studies of neutron stars. To the best of our knowledge, we model here for the first time the recently discovered massive pulsar PSR J0740+6620 and the strangely light HESS compact object via the Tolman IV solution. The present work may be of interest to model builders as well as a useful reference for future research. 
\end{abstract}

\maketitle

%%%%%%%%%%%%%%%%%%%%%%%
\section{Introduction}
%%%%%%%%%%%%%%%%%%%%%%%

General Relativity (GR), formulated by A. Einstein more than 100 years ago \cite{Einstein:1915ca}, is worldwide accepted as a beautiful and at the same time very successful gravitational theory, thanks to which we nowadays understand and describe different aspects of Astrophysics and Cosmology. Many of its remarkable predictions have been confirmed observationally, starting from the classical tests in the old days \cite{Weinberg, MTW}, and more recently with the historical LIGO’s direct detection of gravitational waves emitted from black holes mergers in binaries \cite{Ligo1, Ligo2, Ligo3}. For a recent review on the tests of GR see e.g. \cite{Asmodelle:2017sxn}.

\vskip 0.15cm

Despite the mathematical elegance of GR, the analysis of problems corresponding to realistic physical situations is highly non-trivial in most of the cases of interest, as the field equations are non-linear coupled partial differential equations. The principle of superposition, valid in linear differential equations, does not apply here, and therefore finding exact analytic solutions has been always an interesting and challenging topic, keeping researchers busy for decades. For known exact solutions to Einstein’s field equations, see \cite{ExactSol}.

\vskip 0.15cm

In studies of compact relativistic astrophysical objects the authors usually focus their attention on stars made of an isotropic fluid, where there is a unique pressure along all spatial dimensions, $p_r=p_\theta=p_\phi$. However, celestial bodies are not always made of isotropic fluids only. In fact, under certain conditions matter content of the star may become anisotropic. The review article of Ruderman \cite{aniso1} long time ago mentioned for the first time such a possibility: there the author makes the observation that relativistic particle interactions in very dense media could lead to the generation of anisotropies. In the years to follow, the study on anisotropies in relativistic stars received a boost by the subsequent work of \cite{aniso2}. Indeed, anisotropies may arise in different physical situations of dense matter media, for example phase transitions \cite{aniso3}, pion condensation \cite{aniso4}, or in presence of type 3A super-fluid \cite{aniso5}, to mention just a few.

\vskip 0.15cm

A new and elegant method \cite{Ovalle:2017fgl} which permits us to obtain new solutions starting from a known one has been applied extensively over the years. The so-called Minimal Geometric Deformation (MGD) approach, which was originally introduced in \cite{Ovalle:2007bn} in the context of the brane-world scenario \cite{RS1, RS2}, has been proven to be a powerful tool in the investigation of the properties of self-gravitating objects, such as relativistic stars or black holes \cite{Estrada:2018zbh, Morales:2018urp, Estrada:2018vrl, Ovalle:2018umz}, see also \cite{Ovalle:2008se, Ovalle:2010zc, Casadio:2012pu, Casadio:2012rf, Ovalle:2013xla, Ovalle:2014uwa}. The MGD approach was later extended in \cite{Ovalle:2018gic}, and applied in \cite{Fernandes-Silva:2019fez}, demonstrating the power and potential of that technique. More recently, a method to obtain the isotropic generator of any anisotropic solution was developed in \cite{Contreras:2018gzd}, and applied in \cite{Contreras:2018nfg}.

\vskip 0.15cm

Moreover, recently the concept of complexity for self-gravitating systems within GR was introduced in \cite{herrera}. The so called complexity factor, which is a measure of complexity, appears in the orthogonal splitting of the Riemann tensor. Obviously, it vanishes for homogeneous energy densities and isotropic fluid spheres, but it may also vanish when the two terms containing density inhomogeneity and anisotropic pressure cancel each other, see for instance \cite{comp1, comp2, comp3, comp4, comp5, comp6, comp7, comp8, comp9, comp10, comp11} for works on anisotropic stars within the complexity factor formalism.

\vskip 0.15cm

Yet another approach to obtain exact analytic solutions to the field equations describing hydrostatic equilibrium of relativistic stars is via Karmarkar condition in gravity \cite{karmarkar}, which works well both for isotropic and anisotropic objects. If one of the metric potentials is assumed, then the Karmarkar condition allows us to obtain the other metric potential, and after that the energy density and pressures of the fluid may be computed using the field equations. Within this approach no equation-of-state for the matter content is assumed. For an incomplete list of works obtaining interior solutions via the Karmarkar condition in gravity see e.g. \cite{kar1, kar2, kar3, kar4, kar5, kar6, kar7, kar8, kar9, kar10, kar11}.

\vskip 0.15cm

Regarding the composition and inner structure of compact objects, the most massive pulsars \cite{Demorest, Antoniadis, recent} observed over the last 15 years or so are putting constraints on different equations-of-state, since any mass-to-radius relationship that predicts a highest mass lower than the observed ones must be ruled out. In order to model a certain star, it would be advantageous to know both its mass and its radius, which is not always the case as measuring the radius is way more difficult. There are some good strange quark star candidates, see e.g. Table 5 of \cite{Weber} or Table 1 of \cite{Maxim}, and also the recently discovered massive pulsar PSR J0740+6620 \cite{pulsar1, pulsar2, pulsar3} and the strangely light object HESS  \cite{hess}, where both the stellar mass and radius are known observationally.

\vskip 0.15cm

The exact analytic solution obtained within one of the aforementioned approaches must be physically relevant, namely it must be well behaved and realistic fulfilling all the well established criteria, and it should be capable of modelling one or more of the observed stars of known mass and radius. In this work we propose to model the massive pulsar PSR J0740+6620 \cite{pulsar1, pulsar2, pulsar3} and the strangely light HESS \cite{hess} compact object via the Tolman IV exact analytic solution of GR assuming stars made of isotropic matter. 

\vskip 0.15cm

In the present article our work is organized as follows. After this introduction, in the next section we briefly review the structure equations, while in section 3 we summarize the Tolman IV and the Kohler Chao exact analytic solutions. In the fourth section we model two objects of known mass and radius and check weather or not the corresponding solution is realistic and well behaved. Finally, we conclude our work in the last section. Throughout the manuscript we adopt the mostly positive metric signature, $\{ -, +, +, + \}$, as well as geometrical units, in which both the speed of light in vacuum and Newton's constant are set to unity, $c = 1 = G$.

%%%%%%%%%%%%%%%%%%%%%%%%%%%%%%%%%%%%%%%%%%%%%%%
\section{Anisotropic relativistic stars in GR}
%%%%%%%%%%%%%%%%%%%%%%%%%%%%%%%%%%%%%%%%%%%%%%%

The most general form of a static, spherically symmetric geometry in Schwarzschild-like coordinates, $\{ t, r, \theta, \phi \}$, is given by
\begin{equation}
    d s^2 = - e^{\nu} d t^2 + e^{\lambda} d r^2 + r^2 (d \theta^2 + \sin^2 \theta d \phi^2) 
\end{equation}
where for interior solutions ($0 \leq r \leq R$, $R$ being the radius of the star) $\nu(r), \lambda(r)$ are two independent functions of the radial coordinate. In the discussion to follow it is more convenient to work with the mass function, $m(r)$, defined by
\begin{equation}
    \displaystyle e^{\lambda} = \frac{1}{1 - \frac{2 m}{r}}.
\end{equation}

To obtain interior solutions describing hydrostatic equilibrium of relativistic stars, one needs to integrate the Tolman-Oppenheimer-Volkoff equations \cite{tolman, OV}
\begin{eqnarray}
    m'(r) & = & 4 \pi r^2 \rho (r)   \\ 
    p_r'(r) & = & - [ \rho(r) + p_r(r) ] \; \frac{\nu' (r)}{2} + \frac{2 \Delta}{r} \\
    \nu' (r) & = & 2 \: \frac{m(r) + 4 \pi r^3 p_r(r)}{ r^2 \left( 1 - 2 m(r) / r \right) } 
\end{eqnarray}
where a prime denotes differentiation with respect to $r$, and the matter content is described by a stress-energy tensor of the form
\begin{equation}
T_a^b = Diag(-\rho, p_r, p_t, p_t)
\end{equation}
where $p_r$ is the radial pressure of the fluid, $p_t$ is its tangential pressure, and $\rho$ is the energy density of matter content, while the anisotropic factor is defined by
\begin{equation}
 \Delta = p_t - p_r .
\end{equation}

The function $\nu (r)$ may be computed by
\begin{equation}
    \displaystyle \nu (r) = \ln \left( 1 - \frac{2 M}{R} \right) + 2 \int^r_R \frac{m(x) + 4 \pi x^3 p_r(x)}{ x^2 \left( 1 - 2 m(x) / x \right) } \: dx 
\end{equation}
with $M$ being the mass of the star. The equations (3) - (5) are to be integrated imposing at the centre of the star appropriate conditions 
\begin{equation}
    m(0) = 0 
\end{equation}
\begin{equation}
    p(0) = p_{c}
\end{equation}
where $p_{c}$ is the central pressure. In addition, the following matching conditions must be satisfied at the surface of the object
\begin{equation}
    e^{\nu(R)} = 1-2 M/R
\end{equation}
\begin{equation}
    m(R) = M
\end{equation}
taking into account that the exterior vacuum solution ($r > R$) is given by the Schwarzschild geometry \cite{Schwarzschild:1916uq}
\begin{equation}
    d s^2 = - (1-2M/r) d t^2 + (1-2M/r)^{-1} d r^2 + r^2 (d \theta^2 + \sin^2 \theta d \: \phi^2) .
\end{equation}

%%%%%%%%%%%%%%%%%%%%%%%%%%%%%%%%%%%%%%%%%%%%%%%
\section{Kohler Chao and Tolman IV solutions}
%%%%%%%%%%%%%%%%%%%%%%%%%%%%%%%%%%%%%%%%%%%%%%%

If the metric tensor satisfies the Karmarkar condition \cite{karmarkar}, it can represent an embedding class one spacetime 
\begin{equation}
R_{1414} = \frac{R_{1212} R_{3434} + R_{1224} R_{1334}}{R_{2323}}
\end{equation}
with $R_{2323}$ different than zero. This condition leads to a differential equation given by \cite{kar1}
\begin{equation}
2 \frac{\nu''}{\nu'} +\nu' = \frac{\lambda' e^\lambda}{e^\lambda-1} .
\end{equation}

Upon integration we obtain the relationship between the metric potentials as follows \cite{kar1}
\begin{equation}
e^\nu = \left( A + B \int dr \sqrt{e^\lambda-1} \right)^2
\end{equation}
where $A$ and $B$ are arbitrary constants of integration.

Finally, the anisotropic factor is computed to be \cite{kar1}
\begin{equation}
\Delta = \frac{\nu'}{32 \pi e^\lambda} \left( \frac{2}{r} - \frac{\lambda'}{e^\lambda-1} \right) \left( \frac{\nu' e^\nu}{2 r B^2} - 1 \right) .
\end{equation}
Notice that the anisotropic factor vanishes when at least one of the expressions in the parentheses vanishes. Moreover, it is easy to verify that the $\lambda$ factor being zero corresponds to the Schwarzschild solution. 

In the discussion to follow we shall seek interior solutions describing hydrostatic equilibrium of stars made of isotropic matter. Demanding that the second parenthesis vanishes
\begin{equation}
\frac{\nu' e^\nu}{2 r B^2} - 1 = 0
\end{equation}
then the metric potential, $\nu(r)$, is found to be
\begin{equation}
e^\nu = a + (B r)^2 = a + (r/b)^2
\end{equation}
where the constants $a,b=1/B$ are two free parameters, the first being dimensionless and the other having dimensions of length. Therefore, we recover an existing solution obtained long time ago \cite{Chao}. Thanks to the Karmarkar condition, the second metric potential is computed to be 
\begin{equation}
e^\lambda = 1 + \frac{b^2 (\nu')^2 e^\nu}{4} = 1 + \frac{r^2}{r^2+a b^2}
\end{equation}
while the mass function is given by
\begin{equation}
m(r) = \frac{r^3}{4 r^2 + 2 a b^2}.
\end{equation}
Next, using the structure equations, the pressure and the energy density of the fluid may be computed, and they are given by
\begin{equation}
\rho(r) = \frac{2 r^2+3 a b^2}{8 \pi (2 r^2+a b^2)^2}, \; \; \; \; \; \; \rho_c=\rho(0)=\frac{3}{8 \pi a b^2}
\end{equation}
\begin{equation}
p(r) = \frac{1}{8 \pi (2 r^2+a b^2)}, \; \; \; \; \; \; p_c=p(0)=\frac{1}{8 \pi a b^2} ,
\end{equation}
where the central values $p_c,\rho_c$ are found to be finite. Finally, the speed of sound, $c_s$, and the relativistic adiabatic index, $\Gamma$, are computed to be
\begin{equation}
c_s^2(r) = \frac{dp}{d \rho}=\frac{p'(r)}{\rho'(r)} = \frac{2 r^2+a b^2}{2 r^2+5a b^2}
\end{equation}
\begin{equation}
\Gamma(r) = c_s^2 \left( 1+\frac{\rho}{p} \right) = 4 \frac{r^2+a b^2}{2 r^2+5a b^2} .
\end{equation}

Clearly, neither the energy density nor the pressure become zero at any finite radius, and therefore the Kohler Chao solution is not suitable for describing stellar interior in hydrostatic equilibrium. This is why in the discussion to follow we shall switch to the Tolman IV analytic solution. We comment in passing that since we shall not working with the Kohler Chao solution in the following we could have omitted the discussion altogether. Nevertheless, we have decided to included it in order to demonstrate that modeling stars is not a trivial job, and that not every analytic solution is good enough.

As far as the Tolman IV solution is concerned, we may start assuming a metric potential of the form
\begin{equation}
e^{\lambda(r)}  =  \frac{ 1+2 (r/A)^2 }{[1-(r/r_0)^2] [1+(r/A)^2]}.
\end{equation}
Next, using the definition of the mass function as well the first structure equation, the energy density and the mass function are computed to be
\begin{eqnarray}
\rho(r) & = & \frac{1}{8 \pi r_0^2} \: \frac{6 r^4 + (7 A^2+2 r_0^2) r^2+3 A^4+3 A^2 r_0^2}{(A^2+2 r^2)^2}, \\
m(r) & = & \frac{r^3}{2 r_0^2} \: \frac{r^2+A^2+r_0^2}{A^2+2 r^2}. 
\end{eqnarray}
Finally, making use of the other two structure equations, the pressure of the fluid and the other metric potential, $\nu(r)$, are found to be \cite{Ovalle:2013xla}
\begin{eqnarray}
p(r) & = & \frac{1}{8 \pi r_0^2} \: \frac{r_0^2-A^2-3 r^2}{2 r^2+A^2}, \\
e^{\nu(r)} & = & B^2 [ 1+(r/A)^2 ] .
\end{eqnarray}
As a check, it is straightforward to verify that all structure equations are satisfied. We see that both the pressure and the energy density remain finite at the center of the star, while at the same time the pressure becomes zero at a finite radius $R$, which is found to be 
\begin{equation}
R = \frac{r_0}{\sqrt{3}} \: \sqrt{1-\frac{A^2}{r_0^2}}.   
\end{equation}
Finally, the speed of sound, $c_s$, and the relativistic adiabatic index, $\Gamma$, for the Tolman IV solution are found to be
\begin{equation}
c_s^2(r) = \frac{dp}{d \rho} = \frac{p'(r)}{\rho'(r)} = \frac{2 r^2+A^2}{2 r^2+5 A^2}
\end{equation}
\begin{equation}
\Gamma(r) = c_s^2 \left( 1+\frac{\rho}{p} \right) = 2 \frac{(r^2+A^2) (2 r_0^2+A^2)}{(2 r^2+5 A^2) (r_0^2-A^2-3 r^2)} .
\end{equation}
We notice that contrary to the sound speed that always remains finite throughout the star, the index $\Gamma$ becomes infinite at the surface of the star.

%%%%%%%%%%%%%%%%%%%%%%%%%%%%%%%%%%%%%%%%%%%%%%%%%%%%
\section{Modeling objects of known mass and radius}
%%%%%%%%%%%%%%%%%%%%%%%%%%%%%%%%%%%%%%%%%%%%%%%%%%%%

\subsection{Criteria for realistic solutions}

Before an attempt to model the massive pulsar J0740+6620 and the light object HESS J1731-347 is made, let us first report here the requirements for well-behaved solutions capable of describing realistic astrophysical configurations. Those are the following:

\begin{itemize}

\item Causality, namely the sound speed cannot exceed the speed of light in vacuum
\begin{equation}
0 \leq c_s^2 \leq 1
\end{equation}

\item Stability based on the adiabatic relativistic index, namely its mean value must be larger than a critical value
\begin{equation}
\langle \Gamma \rangle \geq \Gamma_{cr}
\end{equation}
where the critical value is given by \cite{Moustakidis:2016ndw}
\begin{equation}
\Gamma_{cr} = \frac{4}{3} + \frac{19 M}{21 R}
\end{equation}
while the mean value is computed by \cite{Moustakidis:2016ndw}
\begin{equation}
\langle \Gamma \rangle = \frac{\int_0^R dr \: \Gamma(r) p(r) r^2 e^{(\lambda+3 \nu)/2}}{\int_0^R dr \: p(r) r^2 e^{(\lambda+3 \nu)/2}} .
\end{equation}

\item The energy conditions are the constraints on the energy-momentum tensor $T_{\mu \nu}$ of the matter content within a given theory of gravity. The standard acceptable conditions assumed for the energy-momentum tensor are: weak energy condition (WEC), dominant energy condition (DEC), null energy
condition (NEC), and strong energy condition (SEC), see for instance \cite{Ellis, Wald, Frolov}. 
If $\xi_\mu$ and $k_\mu$ are arbitrary time-like and null vectors, respectively, then the conditions for the energy-momentum tensor are expressed with the following inequalities
\begin{equation}      
T^{\mu \nu} \: \xi_\mu \: \xi_\nu \geq 0, \; \; \; \; (WEC)
\end{equation}
\begin{equation}      
T^{\mu \nu} \: \xi_\mu \: \xi_\nu \geq 0 \; \textrm{and} \; T^{\mu \nu} \: \xi_\mu \; \textrm{is a non-spacelike vector,} \; \; \; \; (DEC)
\end{equation}
\begin{equation}      
T^{\mu \nu} \: k_\mu \: k_\nu \geq 0, \; \; \; \; (NEC)
\end{equation}
\begin{equation}      
T^{\mu \nu} \: \xi_\mu \: \xi_\nu - (1/2) T_{\mu}^{\mu} \: \xi^\nu \: \xi_\nu \geq 0 . \; \; \; \; (SEC)
\end{equation}

In particular, in the case of a perfect fluid the energy conditions require that \cite{Panotopoulos:2020uvq, Balart:2023odm}
\begin{equation}      
\rho \geq 0, \; \; \; \; \; \; \rho + 3 p \geq 0, \; \; \; \; \; \; \rho \pm p \geq 0 .
\end{equation}

\end{itemize}

\subsection{Case I: Massive pulsar J0740+6620}

The analytic solution presented in the previous subsection is characterized by three free parameters, $A,r_0$ with dimensions of length, and $B$ which is dimensionless. Their numerical values may be determined imposing the matching conditions 
\begin{equation}
m(R)=M, \; \; \; \; \; \; p(R) = 0, \; \; \; \; \; \; e^{\nu(R)} = 1-2 M/R
\end{equation}
and hence the values of $A,B,r_0$ are found to be
\begin{equation}
A=11.94 km, \; \; \; \; \; \; r_0=24.45 km, \; \; \; \; \; \; B=0.40
\end{equation}
considering stellar mass and radius $M=2.10 M_{\odot}$ and $R=12.32 km$, respectively. Next, we compute the mean value of the adiabatic relativistic index as well as its critical value, which are found to be
\begin{equation}
\Gamma_{cr}=1.56, \; \; \; \; \; \; \langle \Gamma \rangle = 4.05.
\end{equation}
Finally, the metric potentials, the energy density, the pressure, the speed of sound and the adiabatic relativistic index as a function of the radial coordinate are displayed in the Figures \ref{fig:1} and \ref{fig:2} below. We notice that both the pressure and the energy density are finite at the center, and they monotonically decrease throughout the star from the center to the surface. Furthermore, causality is not violated, and all energy conditions are fulfilled as well. Moreover, since $\langle \Gamma \rangle > \Gamma_{cr}$, we conclude that the stability criterion, too, is met by the solution adopted here.

\subsection{Case II: Light object HESS J1731-347}

As in the previous case, imposing the matching conditions we find the following values of the free parameters
\begin{equation}
A=25.07 km, \; \; \; \; \; \; r_0=30.89 km, \; \; \; \; \; \; B=0.81
\end{equation}
considering stellar mass and radius $M=0.80 M_{\odot}$ and $R=10.42 km$, respectively. After that we compute both the mean value of the adiabatic relativistic index and its critical value, and they are found to be
\begin{equation}
\Gamma_{cr}=1.44, \; \; \; \; \; \; \langle  \Gamma   \rangle = 8.36.
\end{equation}
Next, the quantities of interest, namely the metric potentials, the energy density, the pressure, the speed of sound and the adiabatic relativistic index, as a function of the radial coordinate are displayed in the Figures \ref{fig:3} and \ref{fig:4} below. A behavior very similar to the previous case is observed, although the central (normalized) energy density is now three times larger, while at the same time the adiabatic relativistic index takes higher values as we approach the surface of the star. We observe that causality is not violated, and that all the energy conditions are fulfilled as well. In addition to that, both the pressure and the energy density of the fluid are finite at the origin, and they monotonically decrease with $r$. Finally, since $\langle \Gamma \rangle > \Gamma_{cr}$, we conclude that the stability criterion is met as well.

\smallskip

Before we conclude our work a couple of comments are in order. The first one has to do with the matter content of the stars and the underlying equation-of-state. The exact analytic solution adopted here does not assume any equation-of-state, although it may be obtained in parametric form, $\{ \rho(r), p(r) \}$, and in principle it could be studied in a future work. Despite that, given the behavior of the energy density, the sound speed and the relativistic adiabatic index shown in the figures, one may assume that the objects modeled in this work are made either of dark energy based on the Extended Chaplygin Gas (ECG) equation-of-state \cite{Pourhassan:2014cea} or of quark matter in the color-flavor locked (CFL) phase \cite{Alford:2007xm}. The ECG EoS is the following
\begin{equation}
p = - \frac{B^2}{\rho} + A^2 \rho, \; \; \; \; \; A=\sqrt{0.45}, \; \; B=0.0002 / km^2
\end{equation}
while the CFL EoS in parametric form $\{ \rho(\mu), p(\mu) \}$, with $\mu$ being the quark chemical potential, is of the following form  \cite{VasquezFlores:2017uor}
\begin{equation}
    \displaystyle \rho = B + \frac{9 \mu^4}{4 \pi^2} + \frac{9 \alpha \mu^2}{2 \pi^2}
\end{equation}
\begin{equation}
    \displaystyle p = -B + \frac{3 \mu^4}{4 \pi^2} + \frac{9 \alpha \mu^2}{2 \pi^2}
\end{equation}
\noindent where the constant $\alpha$ is given by
\begin{equation}
    \displaystyle \alpha = \frac{2 \Delta^2}{3} - \frac{m_s^2}{6}
\end{equation}
while the numerical values of the parameters are as follows: $B=60 MeV/fm^3, m_s=0, \Delta=100 MeV$. This is one of the 19 viable models (CFL3) shown in Table I of \cite{VasquezFlores:2017uor}. 

Notice that given the form of the equation-of-states above, only the pressure vanishes at the surface of the star, whereas the energy density acquires a non-vanishing surface value.

\begin{figure*}[ht!]
\centering
\includegraphics[scale=1.25]{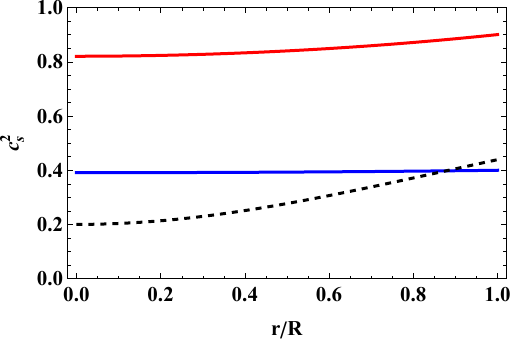} \\
\includegraphics[scale=1.2]{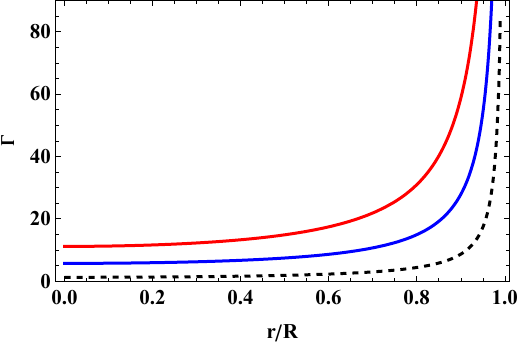}
\caption{
Speed of sound (upper panel) and relativistic adiabatic index (lower panel) versus dimensionless radial coordinate, $r/R$, for the CFL quark matter (curve in blue), the Extended Chaplygin Gas equation-of-state (curve in red) and the Tolman IV solution (black dashed curve). Both panels correspond to the case of the massive pulsar J0740+6620 considering a stellar mass $M=2.10M_{\odot}$.
}
\label{fig:5} 	
\end{figure*}

In Fig. \ref{fig:5} we show the speed of sound as well as the relativistic adiabatic index for both matter contents and for the Tolman solution as well in the case of the massive pulsar J0740+6620. Clearly, the speed of sound of the dark energy content is too large, and so quark matter in the CFL phase is more likely. The behavior of the index $\Gamma$ observed here is the typical behavior whenever the pressure and the energy density cannot be zero at the same time at $r=R$. This may be resolved including a second layer, such as a thin crust, assuming a polytropic equation-of-state, and so both the energy density and the pressure vanish at the surface of the star. Therefore, the relativistic adiabatic index turns out to be finite. This happens for instance in studies of neutron stars.

\smallskip

Finally, in the present work we have assumed stars made of isotropic matter with Einstein's General Relativity. In principle one may consider an alternative theory of gravity. In the latter case, the new field equations will give rise to generalized TOV equations. Therefore, the quantities of interest will be given by different expressions. Moreover, the exterior (vacuum solution) generically speaking will be different than the Schwarzschild geometry. Hence, as a future work it would be interesting to extend this study to other objects of known masses and radii, and also to consider modeling relativistic stars made of anisotropic matter and/or within modified theories of gravity. We hope we can address those issues in forthcoming works in the near future.

%%%%%%%%%%%%%%%%%%%%%%
\section{Conclusions}
%%%%%%%%%%%%%%%%%%%%%%

To summarize our work, we have modeled astronomical objects of known masses and radii within GR via the Tolman IV solution. First, the general description of relativistic stars made of anisotropic matter within Einstein's gravity was reviewed. Next, the so called Kohlar-Chao solution, which was obtained long time ago and which is not suitable for describing stellar interior solutions, was recovered assuming stars made of isotropic matter. After that we presented the Tolman IV solution, and we summarized well established criteria for realistic solutions capable of describing astrophysical configurations. Next, considering the massive pulsar J0740+6620, the three free parameters of the solution were determined imposing the appropriate matching conditions at the surface of the star. Then, the behavior of the solution was displayed graphically. According to our results, the quantities of interest were found to be finite (except for the relativistic adiabatic index that diverges at the surface of the star) and at the same time smooth, continuous functions of the radial coordinate throughout the pulsar. Moreover, well-established criteria, such as energy conditions, causality and stability based on the adiabatic relativistic index, were shown to be fulfilled. Finally, going through the same steps we demonstrated that the light HESS J1731-347 compact object, too, may be successfully modeled, as all the quantities of interest were found to be smooth and continuous functions of the radial coordinate throughout the object, while at the same time all the well established criteria are met as well. The divergence of the index $\Gamma$ may be resolved including a thin crust assuming a polytropic equation-of-state, which is precisely the case seen in studies of neutron stars.

%%%%%%%%%%%%%%%%%%%%%%%%%%
\section{ACKNOWLEDGMENTS}
%%%%%%%%%%%%%%%%%%%%%%%%%%

The author wishes to thank the anonymous reviewers for useful comments and suggestions.

\newpage

%%%%%%%%%%%%%%%%%%%%%%%%%%%%FIGURES%%%%%%%%%%%%%%%%%%%%%%%%%%%

\begin{figure*}[ht!]
\centering
\includegraphics[scale=1]{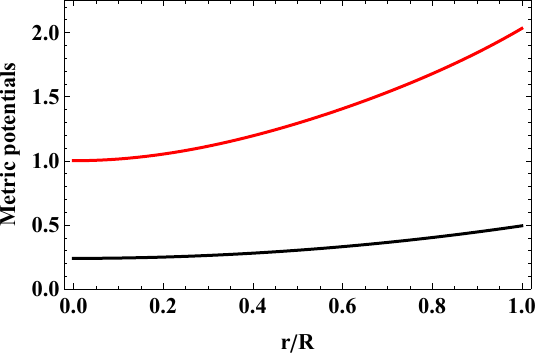} \\
\includegraphics[scale=1.1]{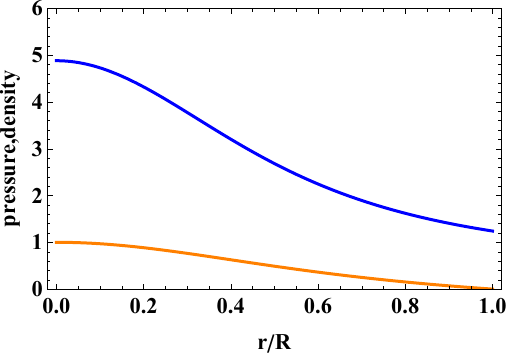} 
\caption{
Metric potentials (upper panel) and normalized pressure and energy density (lower panel) versus dimensionless radial coordinate, $r/R$, for the massive pulsar J0740+6620. The pressure vanishes at $r=R$.
}
\label{fig:1} 	
\end{figure*}

%%%%

\begin{figure*}[ht!]
\centering
\includegraphics[scale=1]{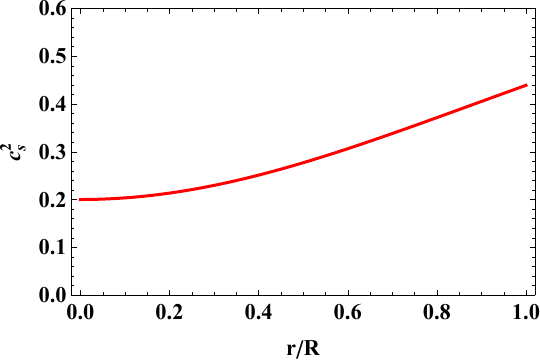} \\
\includegraphics[scale=1]{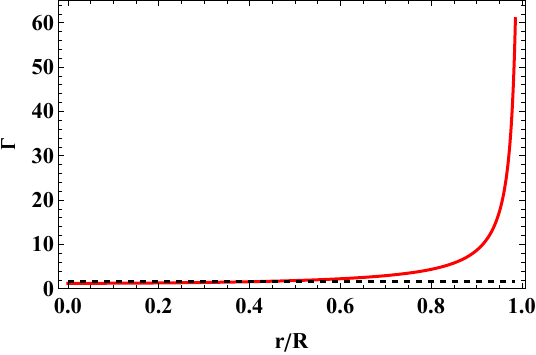}
\caption{
Speed of sound (upper panel) and relativistic adiabatic index (lower panel) versus dimensionless radial coordinate, $r/R$,  for the massive pulsar J0740+6620. The horizontal dashed line corresponds to the critical value.
}
\label{fig:2} 	
\end{figure*}

%%%%

\begin{figure*}[ht!]
\centering
\includegraphics[scale=1]{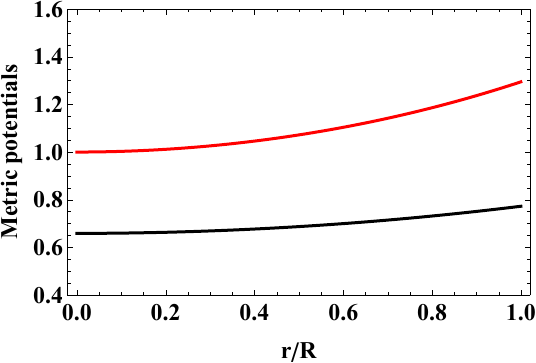} \\
\includegraphics[scale=1.1]{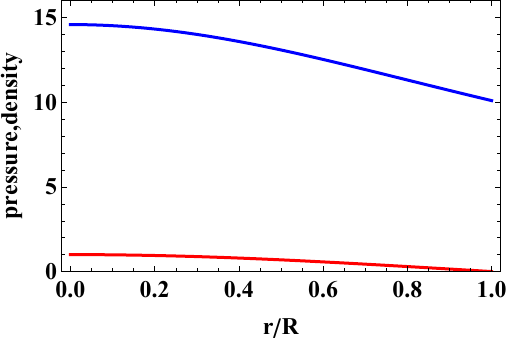} 
\caption{
Metric potentials (upper panel) and normalized pressure and energy density (lower panel) versus dimensionless radial coordinate, $r/R$, for the HESS J1731-347 compact object. The pressure vanishes at $r=R$.
}
\label{fig:3} 	
\end{figure*}

%%%%

\begin{figure*}[ht!]
\centering
\includegraphics[scale=1]{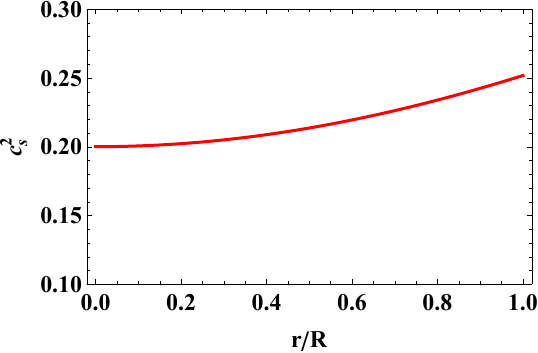} \\
\includegraphics[scale=1]{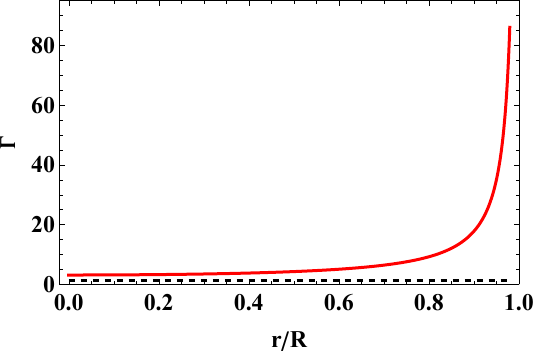}
\caption{
Speed of sound (upper panel) and relativistic adiabatic index (lower panel) versus dimensionless radial coordinate, $r/R$,  for the HESS J1731-347 compact object. The horizontal dashed line corresponds to the critical value.
}
\label{fig:4} 	
\end{figure*}

%%%%%%

%%%%%%%

%%%%%%%%%%%%%%%%%%%END_FIGURES%%%%%%%%%%%%%%%%%%%%%%%%%%%%%%%

\end{document}